\documentstyle[epsfig]{aipproc}

\newcommand{\beq}{\begin{eqnarray}}
\newcommand{\eeq}{\end{eqnarray}}

\newcommand{\la}{\langle}
\newcommand{\ra}{\rangle}
\newcommand{\q}{\la \bar{q}q \ra}
\newcommand{\uq}{\la \bar{u}u \ra}
\newcommand{\dq}{\la \bar{d}d \ra}
\newcommand{\s}{\la \bar{s}s \ra}
\newcommand{\gc}{\la {\alpha_s \over \pi} G^2 \ra}

\newcommand{\nc}{N_c \rightarrow \infty}

\begin{document}

\title{\Large{\bf Multiquark picture for $\Lambda$(1405) and $\Sigma$ (1620)}}
\author{Seungho Choe\thanks{Present address : Dept. of Physical Sciences, Hiroshima University,
Higashi-Hiroshima 739-8526, Japan (e-mail : schoe@hirohe.hepl.hiroshima-u.ac.jp)}}
\address{Department of Physics, Yonsei University,
Seoul 120-749, Korea}
\maketitle
\begin{abstract}
We propose a new QCD sum rule analysis for the $\Lambda$ (1405) and 
the $\Sigma$ (1620).
Using the I=0 and I=1 multiquark sum rules we predict their masses.
\end{abstract}
One of interesting subjects in nuclear physics is to study properties of the
excited baryon states. For example, in the case of the
$\Lambda$ (1405) its nature is not revealed completely\cite{pdg98}; i.e.
an ordinary three quark state or a $\bar{K}N$ bound state or
the mixing state of the previous two possibilities.
In the QCD sum rule approach\cite{qsr} there have been several works 
on the $\Lambda$ (1405) using three-quark interpolating fields\cite{leinweber90,kl97}
or five-quark operators\cite{liu84}.
In this work we focus on the decay modes of the $\Lambda$ (1405) and
the $\Sigma$ (1620) and get the mass of each particle introducing multiquark 
sum rules.

Let's consider the following correlator:
\beq
\Pi (q^2) = i \int d^4x e^{iqx}\langle T ( J(x) \bar{J}(0) )\rangle ,
\eeq
where $J$ is the $\pi\Sigma$ (I=0) multiquark interpolating field,
$J_{\pi^+\Sigma^- + \pi^0\Sigma^0 + \pi^-\Sigma^+}$. 

Here, for the $\Sigma$ we take the Ioffe's choice\cite{ioffe81}; e.g. 
$\pi^0\Sigma^0$ means 
$\epsilon_{abc}(\bar{u}_e i\gamma^5 u_e
			     -\bar{d}_e i\gamma^5 d_e)
([u_a^T C\gamma_\mu s_b]\gamma^5\gamma^\mu d_c
 + [d_a^T C\gamma_\mu s_b]\gamma^5\gamma^\mu u_c) $,
where $u$, $d$ and $s$ are the up, down and strange quark fields, and
$a,b,c,e$ are color indices.
$T$ denotes the transpose in Dirac space and $C$ is the charge
conjugation matrix.

\begin{figure}[t]
\centerline{\epsfig{file=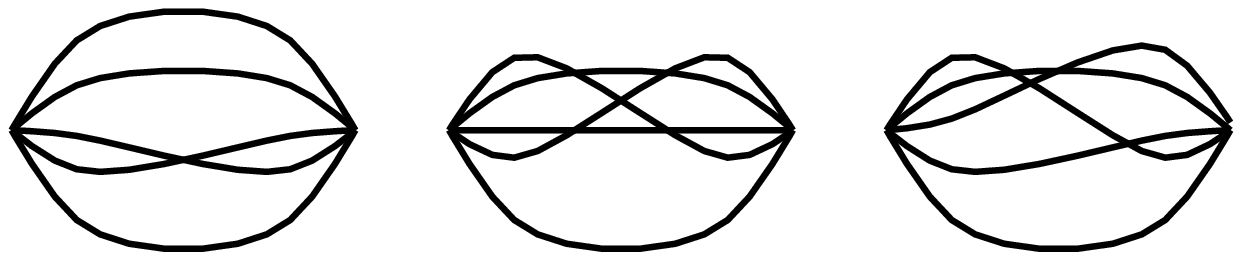,height=1.5cm,width=5cm,angle=0}}
\centerline{\epsfig{file=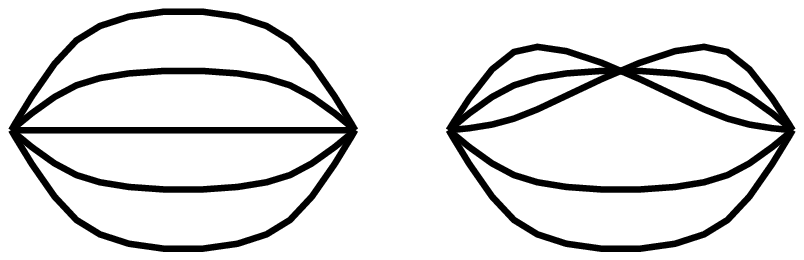,height=1.5cm,width=5cm,angle=0}}
\caption{``bound'' (upper) and ``unbound'' (lower) diagrams.
Solid lines are the quark propagators.}
\label{fig1}
\end{figure}

The OPE side has two structures:
\beq
\Pi^{OPE} (q^2) =\Pi_{q}^{OPE} (q^2) {\bf \rlap{/}{q}}
		+\Pi_{1}^{OPE} (q^2) {\bf 1} .
\eeq
In this paper, however, we only present the sum rule 
from the $\Pi_1$ structure (hereafter referred to as the $\Pi_1$ sum rule)
because the $\Pi_1$ sum rule is generally more reliable than 
the $\Pi_q$ sum rule as emphasized in Ref. \cite{jt97}.
The OPE side is given as follows.
\beq
\Pi_{1}^{OPE} (q^2) = 
 &-& \frac{7 ~m_s}{\pi^8 ~2^{18} ~3^2 ~5} q^{10} ln(-q^2)
+ \frac{7}{\pi^6 ~2^{15} ~3^2} \s q^8 ln(-q^2)
\nonumber\\
&+& \frac{35 ~m_s^2}{\pi^6 ~2^{14} ~3^2} \s q^6 ln(-q^2)
- \frac{121 ~m_s}{\pi^4 ~2^9 ~3^2} \q^2 q^4 ln (-q^2)
\nonumber\\
&+& \frac{11}{\pi^2 ~2^6} \q^2 \s q^2 ln(-q^2)
- \frac{m_s^2}{\pi^2 ~2^6 ~3} (14\q^3 - 33 \q^2\s) ln(-q^2)
\nonumber\\
&-& \frac{m_s}{2^4 ~3^3} (140\q^4 + 3\q^3 \s) \frac{1}{q^2} ,
\label{ope_pisig_i0}
\eeq
where $m_s$ is the strange quark mass and $\q$, $\s$ are the 
quark condensate and the strange quark condensate, respectively.
Here, we let
$m_u$ = $m_d$ = 0 $\neq$ $m_s$ and
$\uq$ = $\dq$ $\equiv$ $\q$ $\neq$ $\s$.
We neglect the contribution of gluon condensates and
concentrate on tree diagrams such as
Fig. \ref{fig1},
and assume the vacuum saturation hypothesis
to calculate quark condensates of higher dimensions.
Note that only some typical diagrams are shown in Fig. \ref{fig1}.

The contribution of the ``bound'' diagrams is a $1/N_c$ correction
to that of the ``unbound'' diagrams, 
where $N_c$ is the number of the colors.
In Eq. (\ref{ope_pisig_i0}) we set $N_c$ = 3.
The ``unbound'' diagrams correspond to a picture that two particles
are flying away without any interaction between them.
In the $\nc$ limit
only the ``unbound'' diagrams contribute to the $\pi\Sigma$
multiquark sum rule.
Then, the $\pi\Sigma$ multiquark mass ($m(\pi\Sigma)$) should be
the sum of the pion and the $\Sigma$ mass in this limit.

\begin{figure}
\centerline{\epsfig{file=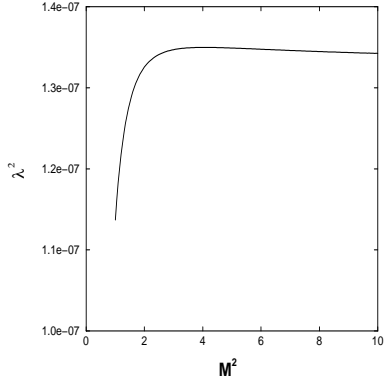,height=5cm,width=5cm,angle=0}}
\caption{The Borel-mass dependence of the coupling strength 
$\lambda^2$ from the $\pi\Sigma$ multiquark sum rule at 
$s_0$ = 2.789 GeV$^2$.}
\label{fig2}
\end{figure}

Eq. (\ref{ope_pisig_i0}) has the following form:
\beq
\Pi^{OPE}_1 (q^2) &=& a ~q^{10} ln(-q^2) + b ~q^8 ln(-q^2)
+ c ~q^6 ln(-q^2) 
+ d ~q^4 ln(-q^2)
\nonumber \\
&+& e ~q^2 ln(-q^2)+ f ~ln(-q^2) + g ~\frac{1}{q^2} ,
\eeq
where $a, b, c, \cdots, g$ are constants.
Then, we parameterize the
phenomenological side as
\beq
\frac{1}{\pi} Im \Pi^{Phen}_{1} (s) &=& \lambda^2 m \delta(s-m^2) +
	     [-a~s^5 - b~s^4 - c~s^3 - d~s^2 - e~s - f] \theta(s~-~s_0) ,
\eeq
where $m$ is the $m(\pi\Sigma)$
and $s_0$ the continuum threshold.
$\lambda$ is the coupling strength of the interpolating field
to the physical $\Lambda$ (1405) state.
The Borel-mass dependence of the $m(\pi\Sigma)$
shows that there is a plateau for the large Borel mass.
However, this is a trivial result 
from our crude model on the phenomenological side.
Hence we do not take this as the $m(\pi\Sigma)$ and neither as the $\Lambda$ (1405) mass.
Instead, we draw the Borel-mass
dependence of the coupling strength
$\lambda^2$ at $s_0$ = 2.789 GeV$^2$ as shown in Fig. \ref{fig2}, where the $s_0$ is taken by
considering the next $\Lambda$ particle\cite{pdg98}.
There is the maximum point in the figure.
It means that the $\pi\Sigma$ multiquark state couples
strongly to the physical $\Lambda$ (1405) state at this point.
Then we take the $\Lambda$ (1405) mass as 
the $m(\pi\Sigma)$ at the point.
However, 
it would be better to determine an effective threshold $s_0$ from 
the present sum rule itself. 

Thus, the steps for getting the $m(\pi\Sigma)$ 
are as follows.
First, consider ``unbound" diagrams only and choose a threshold
$s_0$ in order that the average mass between the fiducial Borel interval 
becomes the $m(\pi)$ + $m(\Sigma)$. Second, consider whole diagrams
(``unbound" + ``bound" diagrams) and draw the Borel-mass
dependence of the coupling strength $\lambda^2$ using the above
$s_0$. Last, determine the $m(\pi\Sigma)$ where the $\lambda^2$
has the maximum value, and thus take this as the $\Lambda$ (1405) mass.
Following the above steps we get the $m(\pi\Sigma)$
= 1.424 GeV at $s_0$ = 3.082 GeV$^2$.

There is another I=0 multiquark state; i.e.
the $\bar{K}^0 n + K^- p$ multiquark state.
Similarly, we obtain the $m(\bar{K}N)$ = 1.589 GeV at $s_0$ = 3.852
GeV$^2$. This corresponds to the $\Lambda$ (1600) mass.
It is interesting to note that the masses 
from two multiquark states are similar at the same threshold
as shown in  Table \ref{table1}.

\begin{table}[b]
\caption{Mass of the $\bar{K}N$, $\pi\Sigma$ (I=0) multiquark states
(~$\q$ = -- (0.230 GeV)$^3$, $\s$ = 0.8~$\q$, and $m_s$ = 0.150 GeV). }
\label{table1}
\begin{tabular}{c c c c c}
  & $s_0$ (GeV$^2$)  &  $m(\bar{K}N)$ (GeV)     &  $m(\pi\Sigma)$ (GeV) & \\
\tableline
  & 3.852          &  1.589         &            1.612    & \\
  & 3.082          &  1.405         &            1.424    & \\
\end{tabular}
\end{table}

Now, we can extend our previous analysis to the I=1 multiquark
states and thus get the $\Sigma$ (1620) mass. 
There are three decay channels for the $\Sigma$ (1620).
Then, we can construct the following multiquark interpolating fields;
$J_{\bar{K}^0 n - K^- p}$, 
$J_{\pi^+\Sigma^- - \pi^-\Sigma^+}$, 
and $J_{\pi^0\Lambda}$ ( or $J_{\pi^\pm\Lambda}$).
In Table \ref{table2} we present each multiquark mass.

\begin{table}[t]
\caption{Mass of  the $\bar{K}N$, $\pi\Sigma$, and $\pi\Lambda$ (I=1)
multiquark states 
(~$\q$ = -- (0.230 GeV)$^3$, $\s$ = 0.8~$\q$, and $m_s$ = 0.150 GeV). }
\label{table2}
\begin{tabular}{c c c c c c}
 & $s_0$ (GeV$^2$)  &  $m(\bar{K}N)$ (GeV)     &  $m(\pi\Sigma)$ (GeV) &  $m(\pi\Lambda)$ (GeV) & \\
\tableline
  & 3.852          &  1.589         &            1.606    &  1.581 & \\
\end{tabular}
\end{table}

We have obtained the I=0 and I=1 multiquark masses which are slightly
different from the experimental values\cite{pdg98}.
One of corrections is
to include the isospin symmetry breaking effects (i.e. $m_u \neq m_d \neq 0$,
$\uq \neq \dq$, and electromagnetic effects) in our sum rules.
On the other hand, one can consider the contractions between 
the $\bar{u}$ and $u$ (or between the $\bar{d}$ and $d$) quarks
in the initial state which have been excluded in our previous calculation.
However, it is found that this correction
is very small comparing to other $1/N_c$ corrections,
i.e. the contribution of ``bound" diagrams.
Another possibility  is the correction from
the possible instanton effects\cite{instanton} to the I=0 and
I=1 states, respectively.    

In this work we have neglected the contribution of gluon condensates
and that of other higher dimensional operators including gluon
components.
Since we have considered the $\Pi_1$ sum rule,
 only the odd
dimensional operators can contribute to the sum rule.
Thus, for example, the contribution of the gluon condensates
is given by the terms like
$m_s \gc $ and thus can be neglected
comparing to other quark condensates of the same dimension.

In summary, the $\Lambda$ (1405) and $\Sigma$ (1620) masses are predicted in the
QCD sum rule approach using
the $\bar{K}N$, $\pi\Sigma$, and $\pi\Lambda$
multiquark interpolating fields (both I=0 and I=1).

The author thanks Prof. D.-P. Min and Prof. C.-R. Ji
for their effort to make NuSS'99 successful. 
This work was supported in part by the Korea Science and Engineering
Foundation (KOSEF).


\end{document}